\documentclass[%
 aip,
 amsmath,amssymb,
 reprint,%
]{revtex4-1}

\usepackage{graphicx}
\usepackage{dcolumn}
\usepackage{bm}

\usepackage[utf8]{inputenc}
\usepackage[T1]{fontenc}
\usepackage{mathptmx}
\usepackage{etoolbox}
\usepackage{float}
\usepackage{epsfig,amssymb,amsmath,latexsym, upgreek}
\usepackage{natbib,hyperref}
\usepackage{graphicx}
\usepackage{dcolumn}
\usepackage{bm}
\usepackage{color}
\usepackage{import}
\usepackage{xr} 
\makeatletter
\def\@email#1#2{%
 \endgroup
 \patchcmd{\titleblock@produce}
  {\frontmatter@RRAPformat}
  {\frontmatter@RRAPformat{\produce@RRAP{*#1\href{mailto:#2}{#2}}}\frontmatter@RRAPformat}
  {}{}
}%
\makeatother

\newcommand{\tF}{t_\text{F}}

\newcommand{\FLDG}{\mathcal{F}_\text{LDG}}
\newcommand{\alphaStar}{\alpha^{*}}
\newcommand{\wQ}{W_\text{Q}}
\newcommand{\wU}{W_\text{u}}
\newcommand{\uT}{\mathbf{u}^\text{T}}
\newcommand{\QT}{\mathbf{Q}^\text{T}}
\newcommand{\Utarget}{U_\text{T}}
\newcommand{\la}{l_\text{a}}
\newcommand{\JT}{\mathcal{J}_\text{T}}
\newcommand{\epsNoise}{\epsilon_\text{n}}
\newcommand{\dxAlpha}{\delta_{\alpha}}

\begin{document}

\preprint{AIP/123-QED}

\title{Achieving designed texture and flows in bulk active nematics using optimal control theory}

\author{Saptorshi Ghosh}
\author{Aparna Baskaran}
 \email{aparna@brandeis.edu}
\author{Michael F. Hagan}%
\email{hagan@brandeis.edu}
 
\affiliation{%
Martin A. Fisher School of Physics, Brandeis University, Waltham, Massachusetts 02453, USA 
}%

\date{\today}

\begin{abstract}
Being intrinsically nonequilibrium, active materials can potentially perform functions that would be thermodynamically forbidden in passive materials. However, active systems have diverse local attractors that correspond to distinct dynamical states, many of which exhibit chaotic turbulent-like dynamics and thus cannot perform work or useful functions. 
Designing such a system to choose a specific dynamical state is a formidable challenge. Motivated by recent advances enabling optogenetic control of experimental active materials, we describe an optimal control theory framework that identifies a spatiotemporal sequence of light-generated activity that drives an active nematic system toward a prescribed dynamical steady-state. Active nematics are unstable to spontaneous defect proliferation and chaotic streaming dynamics in the absence of control. We demonstrate that optimal control theory can compute activity fields that redirect the dynamics into a variety of alternative dynamical programs and functions. This includes dynamically reconfiguring between states, and selecting and stabilizing emergent behaviors that do not correspond to attractors, and are hence unstable in the uncontrolled system.  Our results provide a roadmap to leverage optical control methods to rationally design structure, dynamics, and function in a wide variety of active materials. 
\end{abstract}

\maketitle

\section{\label{sec:level1}Introduction: }

The precise control of fluids is essential for biological processes including motility and material transport across cell, tissue and organismal scales of organization \cite{Fletcher2010, Hawkins2009, Berg2004, Liebchen2018, Pierce2008, Videler1993, Behkam2007, Angelini2011, Duclos2018}. These flows are generated and modulated by `active' protein structures that convert chemical energy into mechanical forces and motion. While these protein sources are generated at the molecular (nm) scale, they self-organize to generate and control fluid flows on much larger (micron) scales. For example, cytoskeletal networks composed of filaments and motor proteins induce cytoplasmic flows or streams within cells that drive processes ranging from chloroplast transport in algae to the recycling of motility proteins during eukaryotic chemotaxis \cite{Allen1978, Keren2009, Monteith2016, Keller1971}. Notably, control and self-regulation are essential to target the cytoskeleton to a particular functional dynamical state, selected from the rich manifold of available states \cite{Blanch2018, Poujade2007, Serra2012}. 
Understanding the mechanisms of biological fluid control will enable developing synthetic systems that use similar principles of self-organization to enable macroscale control of flows, thus creating advanced functional, responsive, and adaptive materials \cite{Shao2014, Qu2021, Wu2017}.

Existing uncontrolled active matter systems composed of purified filaments and motor proteins exhibit nonequilibrium, self-organized structures including microtubule asters \cite{Foster2015, Najma2024, Berezney2022, Hentrich2010, Ndlec1997}, contractile microtubule networks \cite{Foster2015, Najma2024}, and spontaneous fluid flows \cite{Sanchez2012, Decamp2015, Ellis2018, Lushi2014}. However, these flows are typically chaotic and disorganized, and thus not capable of producing work or useful functions. One approach to organize active matter flows is to use confining boundaries \cite{Minu2020, Wu2017, Wioland2013, Lushi2014, Hardouin2019, Opathalage2019, Uplap2023, Fily2014, Fily2015, Fazli2021, Iyer2023}. However, a more versatile approach, typically employed by biological systems, is to implement feedforward and/or feedback control. Experiments have recently taken a key step toward this approach by demonstrating that light can be used to control the spatiotemporal patterns or assembly of active colloids \cite{Aubret2018, Palacci2013} and microswimmers \cite{Volpe2011, Liebchen2018, Huang2020, Winkler2020, Villa2019, Qi2022, Katuri2017, Kurzthaler2024}. Similarly, researchers have recently used optogenetic motor proteins to develop light-activated active nematic systems, enabling controlling the average flow speed \cite{Lemma2023, Zarei2023, Najma2022, Zhang2021}, steering defects \cite{Zhang2021, Repula2024, Zhang2022}, and driving the assembly and motions of asters \cite{Ross2019}. However, despite these remarkable successes, in general it remains unclear how to choose a spatiotemporal sequence of light that will drive the system into a predetermined state with useful functions. This inverse problem is particularly challenging because active matter undergoes an energy cascade from forces generated at the particle scale to mesoscale active stresses that ultimately drive the emergent dynamics. Thus, more robust control protocols are required.

Recently, theoretical works have taken two approaches toward controlling active matter into functionalized states. In the first, investigators subject systems to predetermined activity patterns, and observe the resulting system behaviors to identify possible target state dynamics \cite{Shankar2022, Zhang2021, Nasiri2023, Knezevic2022, Roenning2023, Desgranges2023, Zhang2022}. In the second, researchers choose a predetermined target state, and use techniques such as optimal control theory \cite{Brunton2022, Dullerud2013, Ghosh2024, Norton2020, Shankar2022, Sinigaglia2023, Mowlavi2023, GarciaMillan2024, Grover2018}, optimal transport \cite{Villani2009}, or reinforcement learning \cite{Suri2024, Rotskoff2024, Falk2021} applied to a continuum hydrodynamic model for the active matter system to identify spatiotemporal activity patterns that drive the system toward the target. More broadly, optimal control theory has been used to find control laws for dynamical systems modeled by partial differential equations (PDEs). These efforts include modeling systems ranging from Newtonian fluids \cite{Nabi2017, Passaggia2010} to traffic flows \cite{Gugat2005}, designing optimal interplanetary trajectories \cite{Liu2016}, and identifying optimal trajectories for individual active colloids \cite{Schneider2019, Liebchen2019}.

More closely related to the present article, recent works used optimal control theory to respectively switch the chirality of the steady state in circularly confined active nematics \cite{Norton2020} or alter the direction of or switch between stable attractors (stationary asters and propagating strips) in a polar active fluid \cite{Ghosh2024, Shankar2022}.
 While those results demonstrate the capability to transition between symmetric states that are attractors in the absence of control, here we show that optimal control protocols can be applied much more broadly, driving the system into states that cannot be obtained or stabilized without control, and do not correspond to exact coherent structures \cite{Wagner2022, Wagner2023}. Other works have applied protocols obtained through trial-and-error \cite{Floyd2024}, optimal control solutions \cite{Shankar2022, Norton2020} or symmetry considerations to perform tasks such as steering individual defects \cite{Shankar2024}. Here, we show that optimal control theory can perform significantly more complex tasks, such as eliminating an ensemble of defects and then maintaining a defect-free region indefinitely. The challenging nature of this task is evident from the complexity of the spatiotemporal activity pattern of this control solution, in particular the activity gradients required to prevent defects from entering the boundaries of the square region of interest that we chose. Recently, Serra et al. \cite{Serra2024} showed that one can use optimal control to regulate material transport (e.g. creating material traps) by generating short-time attractors. While that approach is very powerful for specific situations that are globally well-controlled, our approach is straightforward to implement and works well for emergent behavior using local control that is robust to global noise.
 
In this article, we apply optimal control theory to a continuum hydrodynamic model for light-activated active nematics in a bulk (unconfined) system. To build beyond the observations in the previous works, here we show that optimal control theory can drive the system into arbitrary states, including those that would be unstable in the absence of control. In particular, we  demonstrate control solutions that drive the system into very different target states. In the first, we impose a director field configuration ---a flat uniform nematic within a chaotic bulk sample. In the second and third we impose velocity profiles, in which the control protocol drives the system to mimic coherent flow through a `channel' in the absence of any confining boundaries, and then Couette flow without a moving boundary.
 We also show how the control solutions can be analyzed to elucidate the mechanisms that select emergent states in terms of the dynamical equations of the system. 

The layout of the article is as follows. First, we review the hydrodynamic theory and describe the key features of the steady states of an unconfined active nematic. Then, we describe the method for implementing optimal control theory. Next, we present control solutions that drive the system into the target states, and analyze the solutions to interpret the underlying mechanisms. Finally, we conclude with a discussion of potential routes to implement these protocols in experiments, as well as further implications of the results.

\begin{figure}
    \centering
    \includegraphics[width=0.5\textwidth]{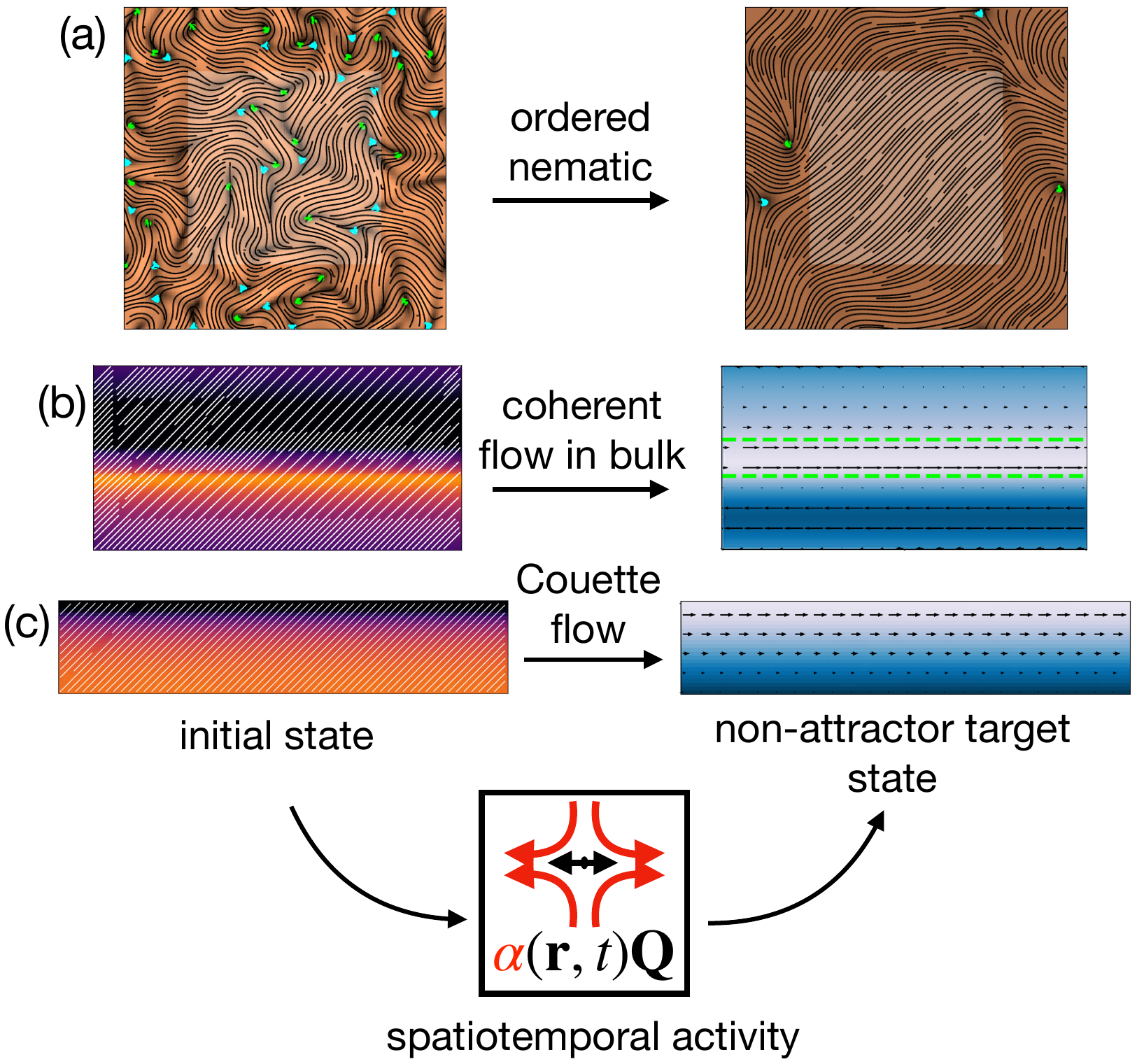}
    \caption{\textbf{Goals for optimal control theory.} 
Starting from a steady state in the absence of control (left images), we identify spatiotemporal activity sequences that drive the system into prescribed states:  \textbf{(a)} a prescribed director configuration, a square domain of uniform active nematic embedded within the chaotic bulk sample; \textbf{(b)} a prescribed velocity profile, coherent flow through a `channel' without confining boundaries; and \textbf{(c)} a prescribed velocity field that mimics Couette flow without a moving boundary. 
}

    \label{fig:introfig}
\end{figure}

\section{\label{sec:model}Model}

As a minimal representation of our system, we use an incompressible, single-fluid model whose state is described by the dimensionless nematic order tensor $\mathbf{Q}=s \rho[\mathbf{n} \otimes \mathbf{n}-(1 / 2) \mathbf{I}]$ and fluid flow field $\mathbf{u}$. $\mathbf{Q}$ describes both the local orientation $\mathbf{n}$ and degree of order $s$ of the nematic, and is scaled by the nematic density $\rho$ such that $\rho s = \sqrt{2 Tr\mathbf{Q^{2}}}$.
 The coupled dynamical equations are given by
 \begin{align}
    \partial_{t} \mathbf{Q}+\nabla \cdot(\mathbf{u} \mathbf{Q})+[\mathbf{Q}, \boldsymbol{\Omega}]-\lambda \mathbf{E}-\mathbf{H} \nonumber\\
    \equiv f_{\mathbf{Q}}(\mathbf{Q}, \mathbf{u}) = 0
    \label{eq:one}
\end{align}
\begin{eqnarray}
    \eta \nabla^{2} \mathbf{u}-\nabla P - \nabla \cdot(\alpha \mathbf{Q}) - \xi\mathbf{u} \equiv f_{\mathbf{u}}(\mathbf{Q}, \mathbf{u}) = 0
    \label{eq:two}
\end{eqnarray}
where $\lambda = 1$ is the flow-aligning parameter, $\eta = 1$ is the coefficient of viscosity, $\xi = 0.01$ is the strength of substrate friction, the mechanical pressure $P$ is such that flows are incompressible: $\nabla \cdot \mathbf{u} = 0 $, and $\Omega_{i j}=\frac{1}{2}\left(\partial_{i} u_{j}-\partial_{j} u_{i}\right)$ and $E_{i j}=\frac{1}{2}\left(\partial_{i} u_{j}+\partial_{j} u_{i}\right)$ are, respectively, the anti-symmetric and symmetric parts of the flow field gradient. We added the small substrate friction to aid convergence of the control solutions; because the associated frictional screening length is much larger than the active length scale for all relevant parameters in this work, it does not qualitatively affect our results {\cite{Doostmohammadi2016b, Alert2022, Chaithanya2024, Schimming2024b, Nejad2021}}. 
The parameter $\alpha$ is the activity ($\alpha > 0:$ extensile, $\alpha < 0: $ contractile).
The \textbf{H} term describes the relaxational dynamics of the nematic tensor to the minimum of a free energy through a molecular field defined as  
\begin{equation}
    \mathbf{H} = - \frac{1}{\gamma}\frac{\partial \FLDG}{\partial \mathbf{Q}}
    \label{eq:H}
\end{equation}

The free energy is given by the Landau–de Gennes free energy functional
\begin{align}
\FLDG=& \int_{\Omega} d^2 \mathbf{r}\left\{\left(-\frac{a_2}{2} \operatorname{Tr}\left(\mathbf{Q}^2\right)+\frac{a_4}{4} \operatorname{Tr}\left(\mathbf{Q}^2\right)^2\right)\right. \nonumber\\ &\left.+ \frac{1}{2} K |\boldsymbol{\nabla} \mathbf{Q}|^2\right\}   .
\label{eq:FLDG}
\end{align}
The dimensionless functions $a_{2} = \rho - 1$ and $a_4 = (\rho + 1)/\rho^{2}$ control the transition from an isotropic fluid ( $\rho < 1$) to a nematic phase ($\rho > 1$); we set the elasticity $K = 2$, rotational viscosity $\gamma = 1$ and $\rho$ =1.6 to focus on the nematic phase far away from the phase transition.

\begin{figure}
  \includegraphics[width=0.5\textwidth]{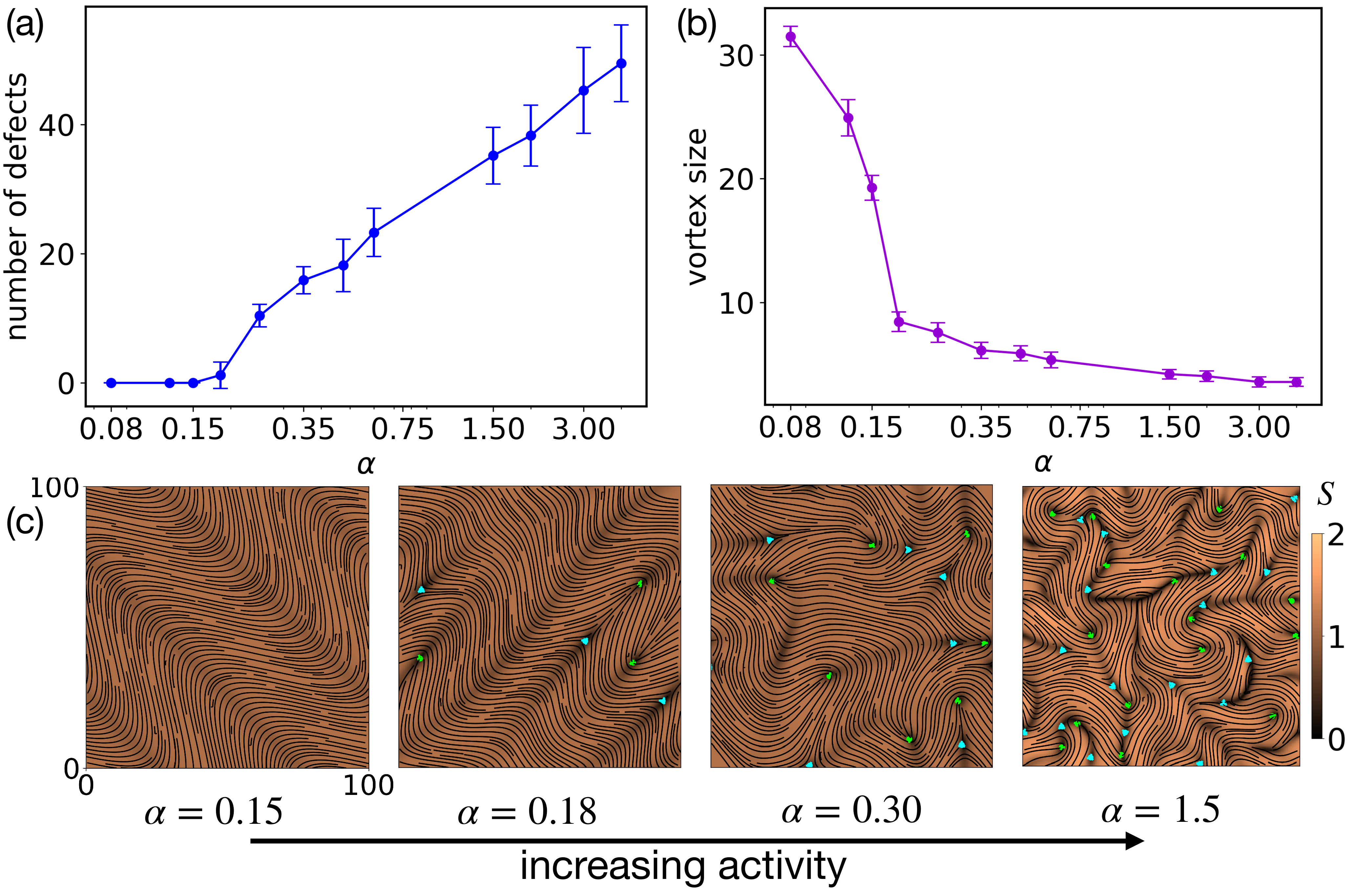}
      \caption{\textbf{Steady-state dynamics of uncontrolled active nematics in bulk with uniform activity. } 
      (a) Number of defects,  (b) mean vortex size, and  (c) snapshots of the steady-state as a function of the activity strength. In (c) the nematic order parameter $s$  and director field orientation are shown by color map and lines respectively. Topological defects are indicated by color, with $+1/2$ in green and $-1/2$ in cyan.  }
      \label{fig:defect_vortex}
\end{figure}  

\begin{figure*}
  \includegraphics[width=\textwidth]{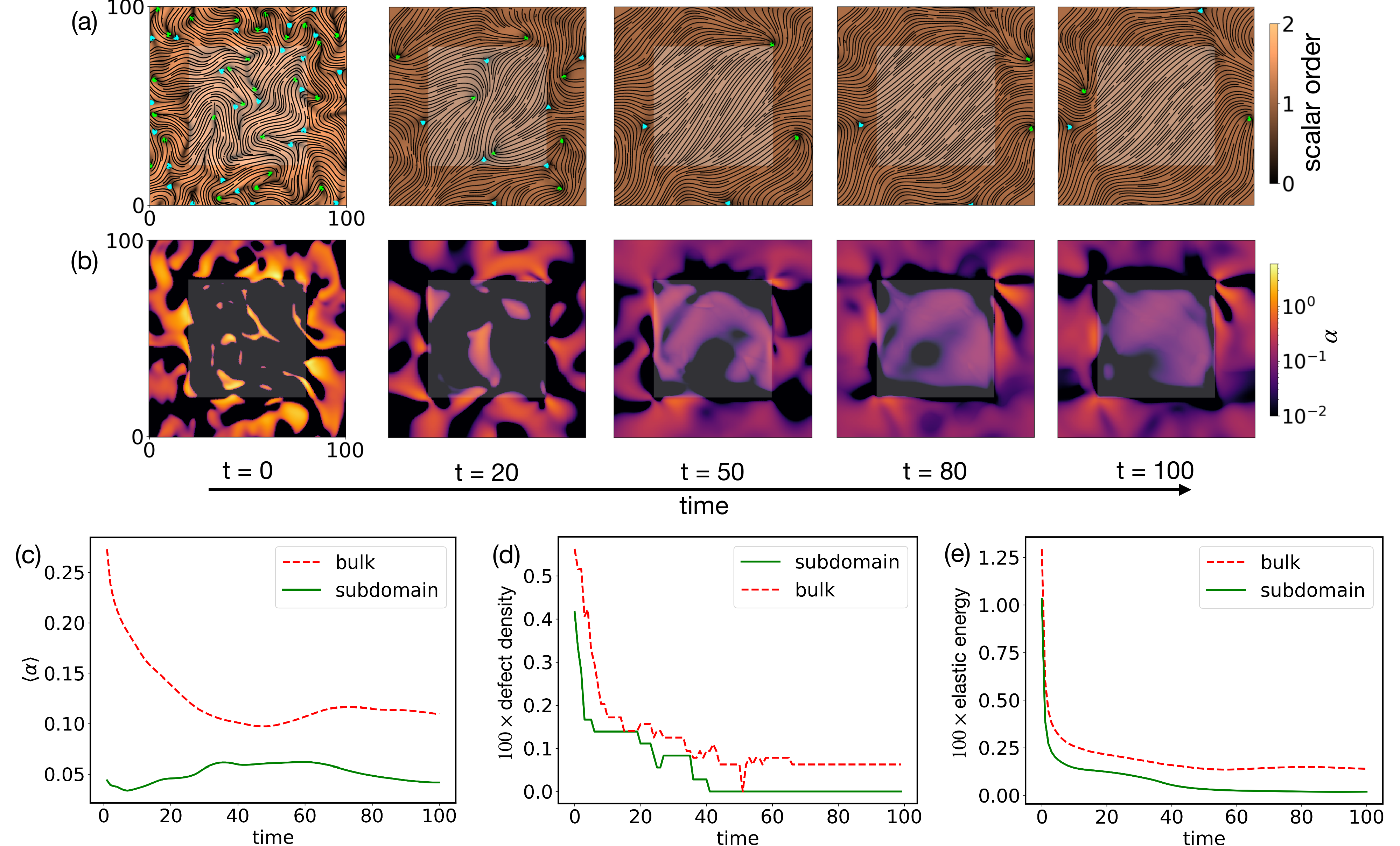}
      \caption{ \textbf{Aligned director field in a square domain.} The target state has a uniform alignment of the director field at an orientation of $\pi/4$ within the square highlighted region. The director field is unconstrained outside of this region, and there are no constraints on the velocity field anywhere. \textbf{(a,b)} Snapshots from a representative trajectory under the optimal control protocol with \textbf{(a)} the nematic order parameter $s$ (color map) and director field orientation (lines)  and \textbf{(b)} the control solution activity field $\alpha$ (color map), with brighter/darker colors representing higher/lower activity. Topological defects are indicated by color, with $+1/2$ in green and $-1/2$ in cyan. 
      The parameters used in this optimization are $\{\alpha_0, J_1, J_2, J_3, \zeta, \wU, \wQ, \tF\} = \{0.0, 1.0, 1.0, 0.1, 0.1, 0.0, 100, 100.0\}$. 
      \textbf{(c,d,e)} Statistics of the system behavior inside and outside the square region:  \textbf{(c)} mean activity $\langle \alpha \rangle$, \textbf{(d)} defect density, and  \textbf{(e)} total elastic energy $\frac{1}{2} K (\mathbf{\nabla Q} )^2$ normalized by spatial extent. A video of this trajectory is in supplemental Movie S1 \cite{SIref}. }
      \label{fig:ordernematic}
\end{figure*}  

\section{\label{sec:optimalControl}Optimal Control}

We search for a spatiotemporal activity field which is extensile ($\alpha(\boldsymbol{x},t) \geq 0$) throughout the control window, and drives and regulates the system toward the target configuration of the nematic tensorial order parameter $\mathbf{Q}(\boldsymbol{x}, t)$ and fluid velocity $\mathbf{u}(\boldsymbol{x}, t)$. We obtain this $\alpha$ by minimizing the scalar objective
functional $\mathcal{J}$, 
\begin{align}
\mathcal{J}=& \frac{1}{2} \int_0^{\tF} \mathrm{~d} t \int_{\Omega} d^2 \mathbf{x}\left[ J_{1}\left(\alpha-\alpha_0\right)^2+ J_{2} \nabla \alpha \cdot \nabla \alpha\right. \nonumber\\
&\left.  + 2 J_{3}\left(-\alpha e^{ -\alpha\zeta}\right) + \wU \Delta \mathbf{u}\cdot \Delta \mathbf{u} + \wQ \Delta \mathbf{Q}: \Delta \mathbf{Q} \right]
\label{eq:J}
\end{align}
subject to Eqs.~\eqref{eq:one} and \eqref{eq:two} 
in the control window, $t \in [0, \tF]$. The term $(\alpha - \alpha_{0})^{2}$ penalises deviations of the magnitude of the control variable $\alpha$, with $\alpha_{0}$ the user-defined baseline activity that controls the mean activity of the system. The term $\boldsymbol {\nabla}\alpha\cdot\boldsymbol{\nabla}\alpha$ favors smoothness of the activity field (to facilitate experimental implementation). The term $(-\alpha e^{-\alpha\zeta })$ penalizes negative values of $\alpha$, which correspond to contractile activities. Further, the terms $\Delta \mathbf{u}$ and $\Delta \mathbf{Q}$ measure the deviation of the state of the system in each time point from its target state, $\uT$ and $\QT$. We constrain our
search of optimal state trajectories to those that comply with the
system dynamics by introducing Lagrange multipliers, $\boldsymbol{\nu}$ and $\boldsymbol{\psi}$, which are adjoint variables for $\mathbf{u}$ and $\mathbf{Q}$, and $\phi$ which enforces incompressibility. These dynamical constraints are enforced in the optimization by adding them to the cost function as 
\begin{align}
\mathcal{L}=\mathcal{J}+\int_0^{\tF} d t \int_{\Omega} d^2 \mathbf{x}\left[\boldsymbol{\nu} \cdot \mathbf{f}_{\mathbf{u}}+\boldsymbol{\psi}: \mathbf{f}_{\mathbf{Q}}+\phi(\nabla \cdot \mathbf{u})\right].
\label{eq:L}
\end{align}

Finally, we solve iteratively for the optimal control variable $\alpha(\boldsymbol{x},t)$  by coupling the Stokes equations with the associated adjoint system.
The conditions for optimality are $(\delta \mathcal{L} / \delta \psi),(\delta \mathcal{L} / \delta \nu)$,
$
(\delta \mathcal{L} / \delta \phi), \quad(\delta \mathcal{L} / \delta \mathbf{Q}), \quad(\delta \mathcal{L} / \delta \mathbf{u}), \quad(\delta \mathcal{L} / \delta P), \\ \quad(\delta \mathcal{L} / \delta \alpha) =0
$. 
The first three conditions return the original nemato-hydrodynamics equations governing 
$\mathbf{Q}$,  $\mathbf{u}$, and $P$. The following three conditions yield the dynamical equations for the adjoint variables $\boldsymbol{\psi}$, $\boldsymbol{\nu}$, and $\phi$  
\begin{align}
    & -\eta\nabla^2 \boldsymbol{\nu}-\nabla \phi+ \xi \boldsymbol{\nu} + \mathbf{h}_1 + \wU(\mathbf{u} - \mathbf{u*})=0 \nonumber \\
    & \nabla \cdot \boldsymbol{\nu} = 0 \nonumber \\
    &\wQ\left(\mathbf{Q}-\mathbf{Q}^*\right)-\partial_t \boldsymbol{\psi}-\mathbf{u} \cdot \nabla \boldsymbol{\psi}-2 \nabla^2 \boldsymbol{\psi} + \alpha \mathbf{h}_2  \nonumber \\
    &\quad+\mathbf{h}_3+2 \mathbf{Q} a_4(\mathbf{Q}: \boldsymbol{\psi})-\boldsymbol{\psi}\left(a_2-a_4 \mathbf{Q}: \mathbf{Q}\right)=0
    \label{eq:adjoint}
\end{align}
with final conditions $\{ \boldsymbol{\psi}, \boldsymbol{\nu} \}(\boldsymbol{x}, \tF) = 0$ where $\boldsymbol{\psi}$ is a symmetric and traceless tensor. The condition ${\delta \mathcal{J}}/{\delta \alpha} = 0$ yields a constraint on the control input as 
\begin{align}
    &J_{1}\left(\alpha-\alpha_0\right)-J_{2} \nabla^2 \alpha + J_{3}\left( \zeta\alpha - 1\right)e^{-\alpha\zeta}\\ \nonumber
    &\quad-\left[Q_{x x}\left(\partial_x \nu_x-\partial_y \nu_y\right)+Q_{x y}\left(\partial_y \nu_x+\partial_x \nu_y\right)\right]=0
    \label{eq:dJ}
\end{align}
Here: $h_{1x} = \psi_{xx}\partial_xQ_{xx} + \psi_{xy}\partial_xQ_{xy} + \partial_y(Q_{xy}\psi_{xx} - Q_{xx}\psi_{xy}) + \lambda(\partial_x\psi_{xx} + \partial_y\psi_{xy})$, 
$h_{1y} = \psi_{xx}\partial_yQ_{xx} + \psi_{xy}\partial_yQ_{xy} + \partial_x(Q_{xx}\psi_{xy} - Q_{xy}\psi_{xx}) + \lambda(\partial_x\psi_{xy} - \partial_{y}\psi_{xx})$, $h_{2x} = \partial_y \nu_y - \partial_x \nu_x$, $h_{2y} = -(\partial_y \nu_x + \partial_x \nu_y)$, $h_{3x} = \psi_{xy}(\partial_y u_x - \partial_x u_y)$ and $h_{3y} = -\psi_{xx}(\partial_x u_y - \partial_y u_x)$. 
 We restrict our search for optimal solutions to the range $\alpha \geq 0$ by calculating the optimal $\alpha(\boldsymbol{x},t)$  via gradient descent, and then actuate 
 \[
\alpha =
\begin{cases}
  0 & \text{if } \alpha < 0 \\
  \alpha & \text{if } \alpha \geq 0
\end{cases}
\]
into the system of coupled differential equations in each step during the optimization. 

We use the direct-adjoint-looping
(DAL) method \cite{kerswell2014} to minimize the cost function under the constraint that the dynamics
satisfies Eqs.~\eqref{eq:one} and \eqref{eq:two}, to yield the optimal schedule of activity in space and time $\alpha(\boldsymbol{x},t)$. 
Specifically, we construct an initial condition corresponding to the stable steady
state at the parameter set of interest, and specify a time duration $\tF$ over which the control
protocol will be employed, and an initial trial control protocol. We then perform a series of DAL
iterations; in each iteration the system  dynamics are integrated from the initial condition $\mathbf{Q}(\boldsymbol{x},0), \mathbf{u}(\boldsymbol{x},0), \alpha(\boldsymbol{x},0)$
for time $\tF$ under the current control protocol, and the cost function \eqref{eq:J} is computed from the
resulting trajectory. The adjoint equations are integrated backwards in time, which propagates the residuals. After each backward run,  the control protocol is updated via gradient descent, $\alpha^{i+i} = \alpha^{i} - \delta \mathcal{L}/\delta \alpha $ using Eq.~\eqref{eq:L}. 
We employ Armijo backtracking \cite{Borzi2011} to compute the step-size of gradient descent and to ensure convergence of the DAL algorithm. We have implemented this calculation in the open source Python finite element method solver FEniCS \cite{Dupont2003}.

\section{Results}

Fig.~\ref{fig:defect_vortex} shows the behavior of the bulk system in the absence of control. We see that for values of activity below a threshold ($\alpha^*\lesssim 0.17 $) the system undergoes small bend fluctuations but does not form defects on simulated timescales. This can be attributed to the long time scale required for defects to spontaneously form at low activity \cite{Giomi2011,Giomi2012,Giomi2013,Marchetti2013,Saintillan2013,Saintillan2018,Theillard2019}. 
Above this threshold, we observe spontaneous defect nucleation, with a density $\propto 1/\la^2 $ and mean vortex size $\propto \la$, with the active length scale $ \la \cong \sqrt{K/\alpha} $ as expected from theory \cite{Marchetti2013}.
Importantly, these states are the only behaviors that we observe above and below the threshold activity value, showing that control is essential to achieve other emerging patterns from this system.

We have used the optimal control implementation described in section~\ref{sec:optimalControl} to compute spatiotemporal activity profiles that drive the system into each of the three target states shown in Fig.~\ref{fig:introfig}. Here we describe the solutions and the associated dynamical trajectories for each case. We also analyze the imposed activity profiles and the system response to gain insight into factors that can stabilize an active nematic director configuration or drive particular flow profiles.

 In each case, we specify the initial and final state, as well as the elapsed time $\tF$ allowed for the control solution. The calculation then yields the activity profile $\alpha(\boldsymbol{x},t)$ and the corresponding system states at every point along the trajectory,  $\mathbf{Q}(\boldsymbol{x},t)$ and $\mathbf{u}(\boldsymbol{x},t)$. Note that all parameters are given in the nondimensional units presented in section~\ref{sec:model}.

\begin{figure*}
  \includegraphics[width=\textwidth]{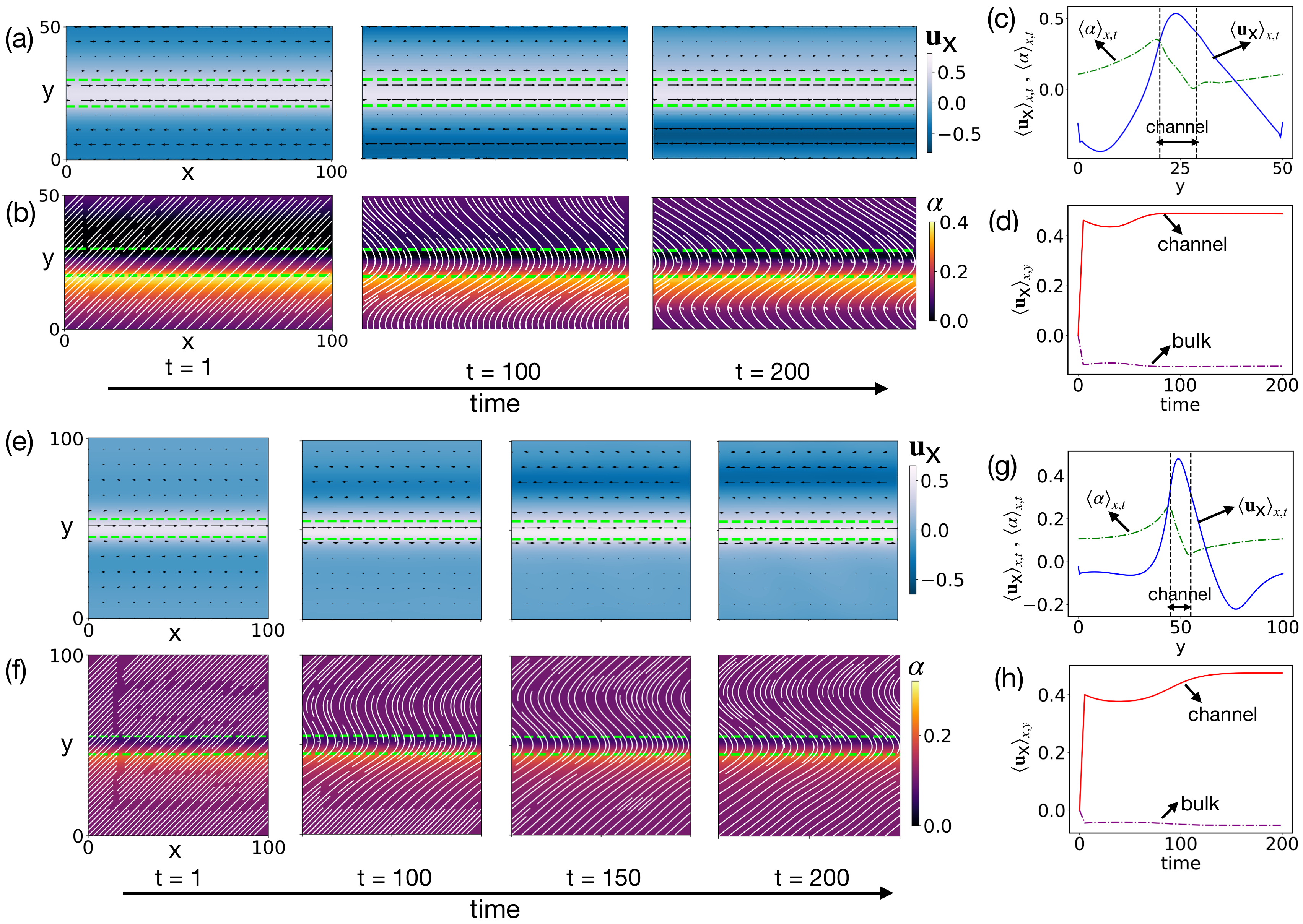}
  \caption{ \textbf{Coherent flow through a `channel' without confining boundaries.} The target state has uniform flow along the $\hat{x}$ direction with magnitude $\Utarget=0.5$ within the strip enclosed by green dashed lines.  \textbf{(a,e)} The velocity field (arrows) and magnitude $u$ (color map) are shown at indicated time points for representative trajectories under the optimal control protocol, for the systems with y-dimension \textbf{(a)} $L_y=50$ and  \textbf{(b)} $L_y= 100 $. \textbf{(b,f)} The activity field $\alpha$ (color map) and nematic director field (lines) are shown at the same time points for the two system sizes. 
 The initial state has the nematic uniformly aligned along the $\pi/4$ axis in each case. 
\textbf{(c,g)} The flow speed $\langle \mathbf{u}_\text{x} \rangle_{x,t}$ and activity $\langle \alpha \rangle_{x,t}$ averaged over time and the channel axis ($\hat{x}$ direction) for the small and large system sizes respectively. \textbf{(d, h)} the flow speed  $\langle \mathbf{u}_\text{x} \rangle_{x,y}$  averaged over the channel and bulk regions as a function of time for the two system sizes.
 The parameter values for this solution were $\{J_1, J_2, J_3, \zeta, \wU, \wQ, \alpha_0, \tF\} = \{1.0, 1.0, 0.1, 0.1, 80.0, 0, 0.1, 200.0\}$.  Videos corresponding to
Videos of these trajectories are provided in supplemental movies S2 and S3 \cite{SIref}. 
}

  \label{fig:coherentflow}
\end{figure*}

\subsection{Nematic alignment of a subdomain within a turbulent bulk active nematic}
\label{sec:orderNematic}

With our first target state, we demonstrate controllability of the nematic tensor field, $\mathbf{Q}$ (Fig.~\ref{fig:ordernematic}). Starting from an initial condition corresponding to a turbulent bulk active nematic, we aim to `flatten' the nematic within a subdomain (highlighted square region in Fig.~\ref{fig:ordernematic}). Specifically, the target state has the nematic uniformly aligned along the $\pi/4$ axis within the subdomain, and no constraints on the director field elsewhere. There are no constraints on the velocity profile $\wU=0$ for this target. To obtain the initial condition, we start with a nearly uniformly aligned nematic (with 1 \% random noise) and zero velocity, and evolve the system with Eqs.~\eqref{eq:one} and \eqref{eq:two} under uniform activity ($\alpha = 2.5$) until it reaches steady state. This results in an initial configuration with 15 defects in the subdomain. We then compute the optimal control solution for $\tF=100$. 

Figs.~\ref{fig:ordernematic}a,b show snapshots of the director field and activity from an example trajectory integrated with the computed optimal activity field. We see that the control solution initially imposes a complex heterogeneous spatiotemporal activity profile within the subdomain to annihilate or drive defects out of the region and reorient the director field along the $\pi/4$ axis. Once this is accomplished, the solution maintains a small (near zero) activity within the subdomain to maintain the flattened state over the remainder of the trajectory. At the same time, the control solution creates a soft confinement to isolate the subdomain from the rest of the sample, which remains chaotic. This is achieved by maintaining gradients of activity at the subdomain boundary that slow and/or reflect approaching defects. For example, at times $t=80$ and $t=100$ in Fig.~\ref{fig:ordernematic}  note the dark region that corresponds to low activity with a high gradient in activity in the vicinity of the subdomain boundary.

\begin{figure*}
  \includegraphics[width=\textwidth]{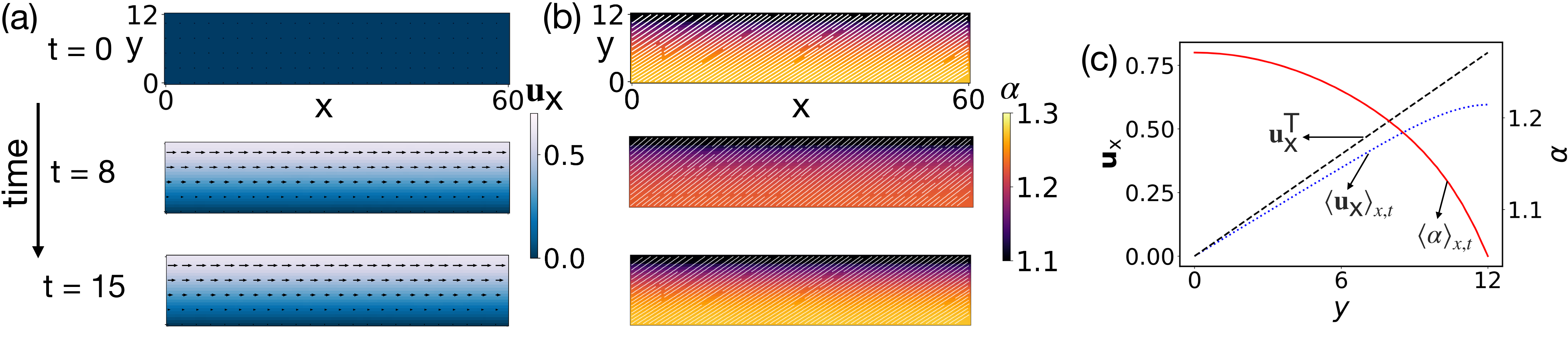}
  \caption{\textbf{Couette flow in a channel without a moving boundary.}  \textbf{(a)} The velocity field (arrows) and magnitude $u$ (color map) are shown for three time points. \textbf{(b)} The control solution activity field $\alpha$ (color map) and nematic director field (lines) for the same time points.\textbf{(c)} Plot of $u_x$ (blue line, left y-axis) compared to the target profile (black dashed line, left y-axis), and the activity (red line, right y-axis) as a function of channel height at the final time point. The parameter values for this solution were $\{J_1, J_2, J_3, \zeta, \wU, \wQ, \tF\} = \{8.0, 1.0, 0.1, 0.1, 15.0, 0, 15.0\}$. A video of this trajectory is in supplemental Movie S4\cite{SIref}.}
  \label{fig:couetteflow}
\end{figure*}

For further insight into the forces that drive flattening of the nematic, we present characteristics of the system behaviors inside and outside of the subdomain as a function of time during the course of the control window (Fig.~\ref{fig:ordernematic}c,d,e). Fig.~\ref{fig:ordernematic}c shows the mean activity as a characteristic of the strength of the control input. We see that the mean activity is relatively constant over time, and significantly smaller inside the subdomain compared to the exterior at earlier times. Over time mean activity of the exterior decreases as well.   We then show the defect density in Fig.\ref{fig:ordernematic}d, average total elastic energy ($1/2 K (\mathbf{\nabla Q} )^2$) in Fig.~\ref{fig:ordernematic}e, and deviation from the target state ($\Delta \mathbf{Q} : \Delta \mathbf{Q} $) in SI \ref{fig:residue}a to characterize the rate at which the solution converges. Interestingly, we see that the elastic energy rapidly decreases both inside and outside the subdomain during the early stage of the trajectory, but reaches a significantly smaller value within the aligned region. The fact that the elastic energy decreases outside of the subdomain, even though this region plays no role in the target state configuration or cost function, can be attributed to the fact that defects are slowed and annihilated as they approach the subdomain, thus decreasing the total defect density and elastic energy of the sample.  This effect can be attributed to the finite size of the sample; in a larger sample there would be a finite correlation length  $\propto \sqrt{K/\langle \alpha\rangle}$  beyond which influence from the subdomain would decay.

\subsection{Coherent flow through a `channel' without confinement}
\label{sec:coherentFlow}
With the next target state, we demonstrate controllability of the velocity field $\mathbf{u}$ of the passive solvent,  through active stresses generated by extensile active nematics. We consider a simulation box with periodic boundary conditions in both directions, with sizes $L_x, L_y$. We then compute an activity profile that first establishes and then maintains flow of a prescribed magnitude along the $\hat{x}$ direction through a `channel', that is a strip through the center of the simulation domain. To ensure that the result is not dependent on any commensurability between preferred wavelengths of bend deformations required to achieve flow and the periodic boundary conditions in the $\hat{y}$ direction, we performed simulations for two system sizes, with $L_x=100$ in both cases, and $L_y= 50 $ and $100 $, respectively.

To allow as much freedom as possible for the solution, the velocity is unconstrained outside the channel and there are no constraints imposed on the director anywhere. Specifically, we set  $\wU=80$ in the channel region and zero elsewhere, and  $\wQ=0$ everywhere. 
Since we have just demonstrated the ability to flatten the director field, we set the initial condition to a uniformly aligned nematic with zero velocity.
We set the baseline activity to $\alpha_0=0.1$ so it is below the threshold for turbulence ($\alphaStar \approx 0.17$). 
We find that achieving this solution takes the controller longer than flattening the director field, and thus we set $\tF=200$.

Figs.~\ref{fig:coherentflow}a,b and \ref{fig:coherentflow}e,f show snapshots of the x-component of the velocity $\mathbf{u}_\text{x}$ and the computed activity field for trajectories integrated with the computed control solutions for the systems with $L_y=50$ and $L_y=100$, respectively. 
To further show how these quantities vary across the channel direction, Figs.~\ref{fig:coherentflow}c,g show the flow speed $\langle \mathbf{u}_\text{x} \rangle_{x,t}$ and activity $\langle \alpha \rangle_{x,t}$ averaged over the  channel axis ($\hat{x}$ direction)  and time for the small and large system sizes respectively. This plot emphasizes that the activity peaks at the lower channel boundary and linearly decreases across the extent of the channel, which drives a peak in the flow velocity at the channel center.
To evaluate convergence,  Figs.~\ref{fig:coherentflow}d,h show the flow speed  averaged over the channel region and exterior region as a function of time. We see that the fluid in the exterior becomes nearly quiescent at very early times, but the flow speed in the channel does not saturate until $t >100$. To present more information about convergence,
 the cost function residuals are shown as a function of time in SI Fig. \ref{fig:residue}b.

The control solution drives the intrinsic active nematic bend instability  \cite{Thampi2014, Giomi2014, Marchetti2013}, where a small bend fluctuation in the director field is reinforced by local shear, resulting in a global bend configuration and coherent flow. This configuration is closely analogous to the well-known configuration of coherent flow in a channel \cite{Wagner2022, Shendruk2017}, but here it is stabilized by the applied activity field instead of boundary anchoring. In particular, we see that the solution imposes uniform activity along the long axis of the subdomain $\hat{x}$, but spatial variation along the orthogonal $\hat{y}$ direction that drives and then stabilizes the bend configuration. Notice that for both system sizes the activity is focused in the lower half of the subdomain, but stabilizes a bend configuration and hence flow along the center of the subdomain. Also, the bend deformation has a similar wavelength in both system sizes, commensurate with the channel width, but the larger system size adopts a large region of uniform nematic below the channel.

The target states also resemble banded states and vortex lattices that are observed in systems with significant substrate friction \cite{Caballero2023, Doostmohammadi2018, Doostmohammadi2016, Schimming2024}. However, those states are highly symmetric, with the width and number of bands determined by confinement and defect spacing. In contrast, our control solution determines the width and number of bands (in this case one) independent of the natural defect spacing.

To understand the limits of controllability, we also performed the optimal control calculation with the baseline activity set to $\alpha_0=0.3$,  above the threshold for turbulence. We were unable to find a stable solution in this case. Interestingly, the controller initially achieves coherent flow in the channel, but with a very different profile. SI Fig.~\ref{fig:flowhigh} a shows snapshots from the resulting trajectory. We see that the system establishes a set of counterrotating vortices that drive flow through the channel region. However, due to the turbulent nature of the system at this activity value, this configuration becomes destabilized by the formation and motions of additional defects, and the magnitude of flow degrades. Although the controller tries to restore the solution, it is able to do so only transiently.

\begin{figure*}
  \includegraphics[width=\textwidth]{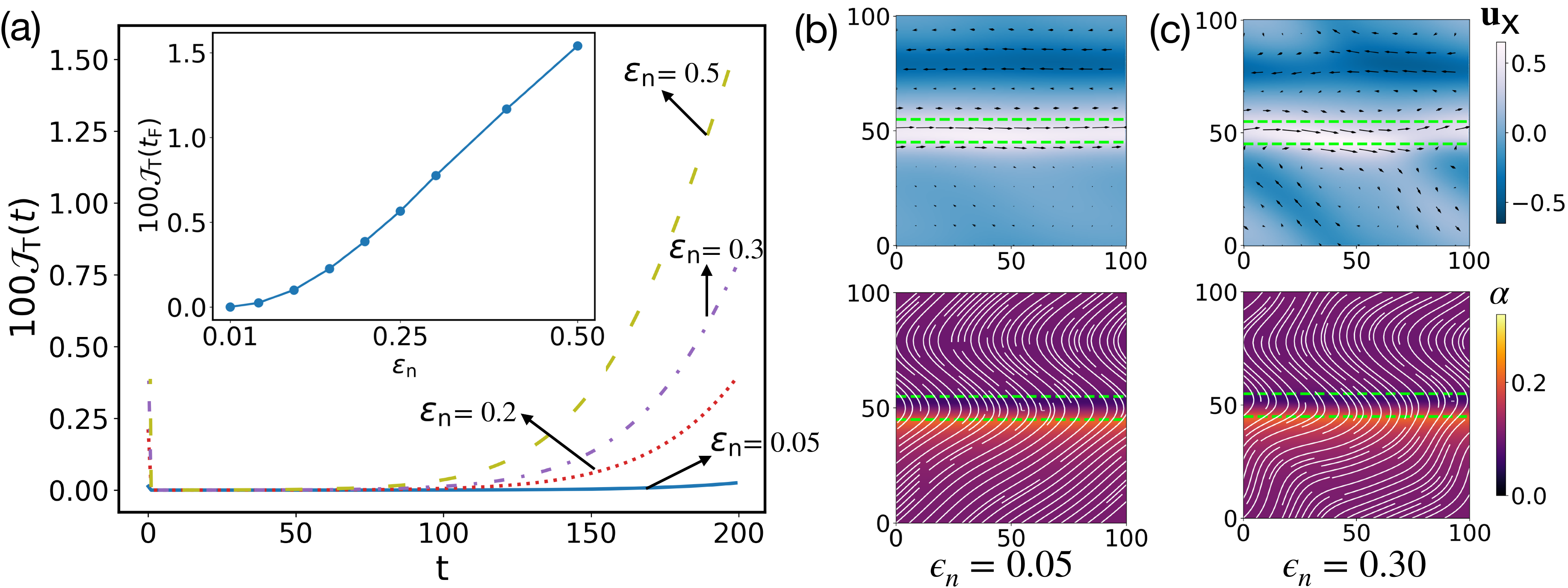}
  \caption{\textbf{Robustness of the control solution to adding noise to the initial condition.}  \textbf{(a)} Deviation as a function of time between the trajectories integrated computed with and without noise ($\JT$, Eq.~\eqref{eq:JT}). We added Gaussian noise to the initial condition with indicated magnitude $\epsNoise$, and integrated the dynamics using the control protocol computed in the absence of noise ($\epsNoise=0$) that is shown in Fig.~\ref{fig:coherentflow}. The inset shows the deviation of the final state $\JT(\tF)$ as a function of noise magnitude $\epsNoise$. \textbf{(b,c)} The final states for \textbf{(b)} small ($\epsNoise = 0.05$) and \textbf{(c)} large ($\epsNoise=0.30$) noise. We see robust flow patterns even when the error is large in the end state for 30\% noise.
}
  \label{fig:noise}
\end{figure*}
\subsection{Couette flow in a channel}
We further demonstrate the controllabilty of velocity field by computing an activity field that causes the system to attain a Couette flow structure in a channel that has no moving boundary. We consider a channel of width $h$, with periodic boundary conditions along the $\hat{x}$ direction, a rigid boundary at the lower wall $y=0$, and a free boundary at the upper wall $y=h$. To mimic Couette flow, we 
set our target velocity profile to $\uT_x = \Utarget y/h$, so that the x-component linearly increases from zero at the lower wall to a maximum $\Utarget$ at the upper wall. We impose no constraints on $u_y$ or $\mathbf{Q}$.  Other boundary conditions are as follows. All components of $\mathbf{u}$ and $\mathbf{Q}$ are periodic in the  $\hat{x}$ direction. For the lower channel wall ($y=0$), we impose no-slip conditions on the velocity, $u_x=u_y=0$. For the upper wall ($y=h$) we set $u_y=0$, but only no-flux (Neumann) conditions on the x-component, $\partial_y u_x=0$. We set Neumann conditions on $\mathbf{Q}$ at both walls. The dimensions of the simulation box are $60 \times h$ with $h=12$. Finally, our initial condition is a uniformly aligned nematic and zero flow velocity.
Figs.~\ref{fig:couetteflow}a,b show the velocity and activity fields at three time points over the course of the solution trajectory.
 Along the channel we find that the activity remains constant, whereas across the channel the activity profile can be computed analytically by solving the Stokes equation \ref{eq:two} with the assumption that Couette flow is established and the director field is initialized at $\theta  = \pi/4$. Neglecting pressure gradients, the Stokes equation simplifies to, $\xi u_x(x, y) = -\partial_{y} ( \alpha(x, y) Q_{xy} )$ and $\partial_x \alpha (x, y) = 0$,
which can be simplified as $\alpha(x, y) \sim \Tilde{\alpha}_0 -\xi\Utarget y^{2}/h $,
with maximum activity at $y = 0$ and minimum  at $y = h$  (Fig.~\ref{fig:couetteflow}c), with $\Tilde{\alpha_0} > 0$. This activity profile mimics the diffusion of momentum from a boundary that would stabilize Couette flow in the canonical moving boundary system. Interestingly, the velocity magnitude deviates below the target near the upper wall. We believe that this residual reflects the fact that a completely linear profile is incompatible with the geometry and boundary conditions that we have imposed. We found that the residual was insensitive to changing controller parameters, such as the relative magnitudes of $J_1, J_2,$ and $\wU$.  
Finally, we note that we obtain stable solutions only for values of $\Utarget<1.0$; higher values give rise to nucleation of defects which can disrupt the steady flow structure and thus impede convergence.
\begin{figure*}
  \includegraphics[width=\textwidth]{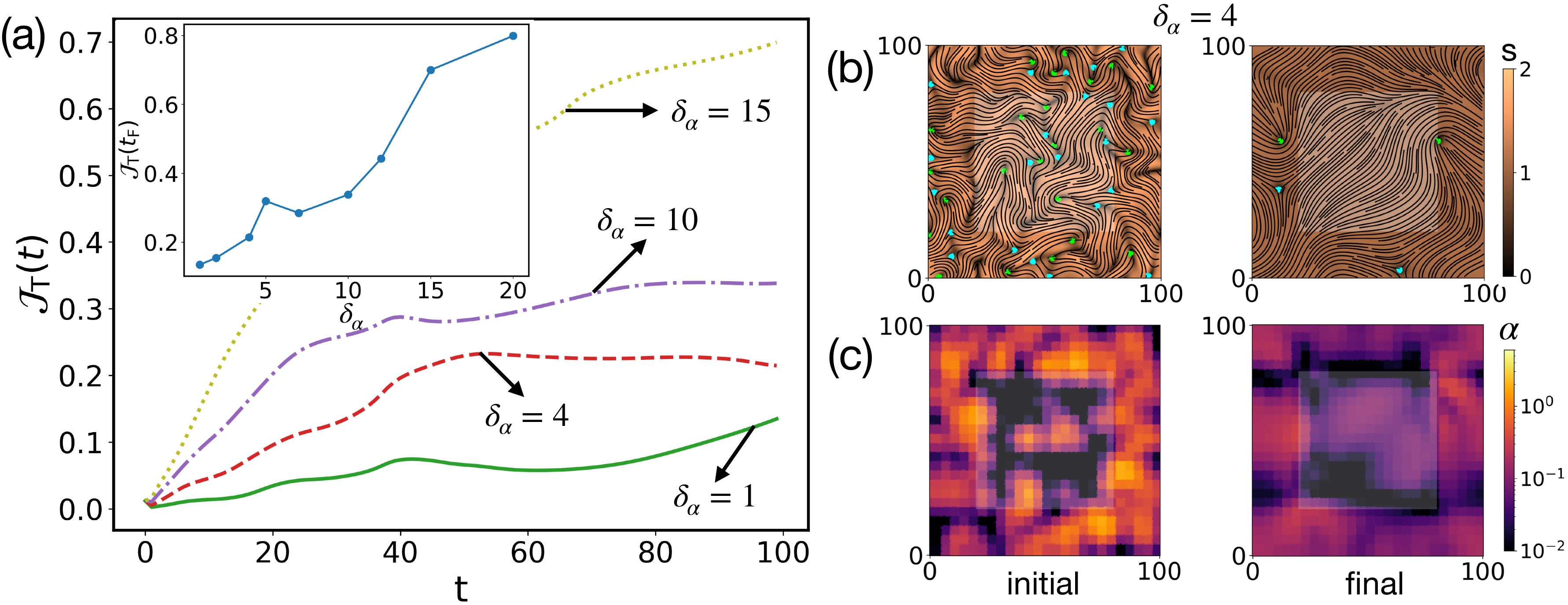}
  \caption{\textbf{Robustness of the control solution to the coarseness of control field}  \textbf{(a)} Deviation as a function of time between the trajectories integrated computed with coarser and optimal $\alpha$ ($\JT$, Eq.~\eqref{eq:JT}). Fig.~\ref{fig:coherentflow}. The inset shows the deviation of the final state $\JT(\tF)$ as a function $\delta \alpha$. \textbf{(b,c)} The initial and final states and activity profile for $\delta_\alpha = 4$. We see low errors for $\delta_\alpha \lesssim 12$, which is the defect spacing at these parameters. 
}
  \label{fig:coarsecontrol}
\end{figure*}

\subsection{Robustness of the optimal control solution to perturbations.} While the optimal control solution in this work is computed for a deterministic PDE model, the ultimate goal is to apply that solution to an experimental system. Thus, it is important to understand how robust the computed solution is to perturbations, since models are not completely accurate and experiments inevitably have noise and measurement errors.

 In previous work \cite{Ghosh2024}, we found that the optimal control solution to an active polar fluid was robust to noise. We observe the same robustness in this active nematic system. In particular, we focus on the coherent flow solution shown in Fig.~\ref{fig:coherentflow}. We perturb the initial condition by adding Gaussian noise with magnitude $\epsNoise$ to the values of $ \mathbf{Q} $ and $\mathbf{u}$ at each lattice site, and then integrate the forward solution of the dynamics with the optimal control solution that we computed without noise. As a measure of error, Fig.~\ref{fig:noise} shows the deviation of the trajectory in the presence of noise compared to the original trajectory without noise 
\ref{fig:coherentflow}  
\begin{align}
\JT(t) = & \frac{1}{\Omega} \int_{\Omega}d\mathbf{r} \frac{1}{2} \left[ (\mathbf{Q}(\mathbf{r},t) - \mathbf{Q}^{*}(\mathbf{r},t))^{2} \right. + \nonumber \\
& \left. (\mathbf{u}(\mathbf{r},t) - \mathbf{u}^{*}(\mathbf{r},t))^{2} \right], 
\label{eq:JT}
\end{align}
where  $(\mathbf{Q}^{*}(\mathbf{r},t), \mathbf{u}^{*}(\mathbf{r},t))$ is the trajectory in Fig.~\ref{fig:coherentflow} and the error is normalized by the spatial extent $\Omega$.
The observation that we obtain consistent trajectory even for large perturbations to the initial condition suggests that the optimal control solution is robust to noise. However, we consider approaches for cases in which the solution becomes inaccurate due to errors in section~\ref{sec:discussion}. Importantly, at all noise values deviations remain extremely small until about the middle of the trajectory, $t = 100$, indicating that the error accumulates over time for it affects the solution. This trend suggests that the feedback approaches  to address noise or errors discussed in section~\ref{sec:discussion} can be applied successfully. 

We also investigated the sensitivity of the control solution to the spatial resolution of the applied light field, since only limited resolution will be available in experimental systems (see section~\ref{sec:discussion} for a discussion of the available resolution in the experimental light activated nematic system). In this case, we use the flattening of the domain solution shown in Fig.~\ref{fig:coarsecontrol}. We coarse-grain the optimal activity $\alpha$ onto a coarser grid with  ${L_x}/{\dxAlpha} \times {L_y}/{\dxAlpha} $ grid points, where $\dxAlpha$ is the coarse-graining parameter. The lower spatial resolution $\alpha$ is smoothed by applying a multidimensional Gaussian filter. The filter computes a weighted sum of neighboring points based on a Gaussian kernel, where the weights are determined by the Gaussian function $G(x) = {1}/{\sqrt{2\pi\sigma^2}} e^{-\frac{x^2}{2\sigma^2}}$. The Gaussian filter is applied independently along each axis, with the 2D filter expressed as: $\alpha_{\dxAlpha} = \sum_{i=-K}^{K} \sum_{j=-K}^{K} G(i, j) \alpha_{\text{coarse}}$
where $G(i,j)$ is the Gaussian kernel. The parameter $\sigma = 0.7$ and kernel size $K = 4$ controls the spread of the filter \cite{Scipy}. 
We then integrate the forward solution using the interpolated activity field. Fig.~\ref{fig:coarsecontrol} shows the deviation from the trajectory computed with the original resolution (Eq.~\eqref{eq:JT}) as a function of $\dxAlpha$. We see that deviations from the original trajectory remain relatively small until the resolution becomes on the order of the defect spacing, $\dxAlpha\approx12$.   This threshold makes sense, as it sets the characteristic length-scale for gradients in the nematic texture that control collective defect motions (as opposed to the shorter-scale gradients at the core of a defect). Moreover, the solutions still achieve the objective (uniform nematic in the region of interest) reasonably well (Fig.~\ref{fig:coarsecontrol}b, c).

Finally, if one knows the experimental limit on resolution of the applied field a priori, one can compute the optimal control solution with the activity field constrained to that resolution. That currently is not possible in our implementation, since the resolution of the activity field is fixed to the finite element grid, but it would be straightforward to include this feature in the framework.

\section{Discussion and conclusions}
\label{sec:discussion}

We have described a framework that employs optimal control theory to drive active nematics into arbitrary emergent behaviors, including states that are not basins of attraction without control. We demonstrate the ability to prescribe both the director and velocity fields. Starting in the turbulent regime, we obtain a uniform nematic aligned along a chosen direction within a prescribed region. Then, we drive the system into states that mimic coherent flow through a channel or Couette flow (Fig. 5A) without the presence of any confining boundaries. Notably, the latter are states that can produce work and perform useful functions, such as powering microfluidics devices. The uniform nematic state also has desirable applications, and furthermore can serve as an initial condition from which other states are readily accessible. These results are also notable for their complexity. For example, rather than steering an individual defect, the flattened domain required eliminating an ensemble of defects and then maintaining a defect-free region indefinitely. The challenging nature of this task is evident from the complexity of the spatiotemporal activity pattern of this control solution, in particular the activity gradients required to prevent defects from entering the boundaries of the square region of interest that we chose.

We also identify limitations on states that can be stabilized --- when the preferred (baseline) activity $\alpha_0$ is set above a threshold, the channel-like flow state becomes unstable to formation of defects and corresponding counterrotating vortices. This reduces the net flow velocity, causing the control solution to repeatedly drive the system back toward stable flow. Finally, we elucidate the mechanisms that drive formation of emergent states and stabilize them by analyzing the activity patterns of the optimal control solutions.

The optimal control protocols can be implemented in any experimental system that allows external actuation of the activity (or other system parameters). For example, researchers have recently constructed light-activated microtubule-based nematics using optogenetic kinesins molecular motors \cite{Lemma2023,Zarei2023,Zhang2021,Ross2019}. When light is shone on these motors, they bind into clusters capable of driving relative displacements of microtubules that power the active nematic; in the absence of light they do not bind and hence are inactive. Lemma et al. and Zarei et al. \cite{Lemma2023,Zarei2023} showed that the magnitude of activity is proportional to applied light intensity, and used digital light projectors to apply spatiotemporal sequences of light. Thus, the optimal control solutions obtained through our framework can be directly implemented in the light-activated nematic system as an open-loop feed-forward control protocol. Importantly, if we map the nondimensional length scale of our equations onto the light-activated nematic system (by matching the mean distance between defects), we find that our results can be obtained with applied activity fields that have spatial resolutions on the order of the experimental capability. In particular, the current experimental spatial resolution on the applied light is 70 $\upmu$m, compared to a typical defect spacing of 250 $\upmu$m\cite{Zarei2023}. As shown in Fig.~\ref{fig:coarsecontrol}, the computed activity field can be interpolated to coarser scales as large as the typical defect spacing ($\approx 12$  non-dimensional length units), suggesting that the experimental resolution is more than sufficient. 

Our previous work has shown that optimal control solutions are robust to noise, and thus can accommodate inaccuracies in the hydrodynamic theory of the active nematic system and experimental noise. If these sources of error are too large, it is straightforward to correct optimal control solutions by using feedback control. For example, one can periodically observe the state of the system along the course of the trajectory, and then either recompute the optimal control solution from the current state of the system or add a linear restoring force that drives the system to return to the computed trajectory \cite{Bechhoefer2021}. As shown in Fig.~\ref{fig:noise}, errors do not accumulate significantly until timescales on order of half the trajectory, suggesting that feedback would not have to be implemented too frequently.

Although we have focused on active nematics in this work, the optimal control framework described here can be easily adapted to any system that can be externally actuated and for which a continuum model description can be obtained. 
We consider actuation of the activity field since that can be directly implemented in the light-activated nematic system \cite{Lemma2023, Zarei2023,Zhang2021,Ross2019}, but the objective function can be defined in terms of any variable that can be externally actuated. For systems that do not already have continuum model descriptions, recent work has demonstrated success in using data-driven model discovery tools to derive PDE descriptions of active matter \cite{Joshi2022, Dunkel2023, Golden2023} and other systems \cite{Fuhg2024, Linden2022, Ramezanian2022}. Furthermore, for systems not amenable to continuum theory, we have recently shown that optimal control solutions can be efficiently calculated for particle-based simulation models by constructing Markov state models \cite{Trubiano2022, Trubiano2024}. Thus, the optimal control framework can be applied to a wide variety of active and other soft material systems, both to achieve complex, functional states and to gain physical insights into the mechanisms that drive and stabilize such emergent dynamics.

\begin{acknowledgments}
This work was primarily supported by the Department of Energy (DOE) DE-SC0022291. Preliminary simulations of the system without control were supported by the National Science Foundation (NSF) DMR-1855914 and the Brandeis Center for Bioinspired Soft Materials, an NSF MRSEC (DMR-2011846). Computing resources were provided by the NSF XSEDE allocation TG-MCB090163 (Stampede and Expanse); the National Energy Research Scientific Computing Center (NERSC), a Department of Energy Office of Science User Facility using NERSC award BES-ERCAP0026774; and the Brandeis HPCC which is partially supported by the NSF through DMR-MRSEC 2011846 and OAC-1920147. AB acknowledges support from NSF - 2202353 and the hospitality of the Aspen Center for Physics, which is supported by National Science Foundation grant PHY-2210452. SG would like thank Fernando Caballero, Michael M. Norton, Chaitanya Joshi and Suraj Shankar for helpful discussions. 
\end{acknowledgments}

\appendix

\section{Direct-adjoint-looping (DAL) method}
\label{sec:DAL}
 We use direct-adjoint-looping (DAL), an iterative optimization method \cite{Kenway2019} to solve for the optimal schedule of activity in space and time that accomplishes our control goals. We start by writing the Lagrangian $\mathcal{L}$ of optimization, Eq. \eqref{eq:L}, where $\boldsymbol{\nu}$ and $\boldsymbol{\psi}$ act as Lagrange multipliers or adjoint variables that constrain the dynamics to follow Eqs.~\eqref{eq:one} and \eqref{eq:two}.

 We construct an initial condition by performing a simulation with unperturbed dynamics (Eqs.~\eqref{eq:one} and \eqref{eq:two}) until reaching steady-state, at a parameter set that leads to a desired initial behavior. Since our target states are not steady states, we define the target configuration by explicitly specifying the desired state variables. We also specify a time duration $\tF$ over which the control protocol will be employed, and an initial trial control protocol $\alpha^{0}(\mathbf{r}, t)$. 
We then perform a series of DAL
iterations, with each iteration involving the following steps:
\begin{itemize}
    \item \textbf{Step 1:} The equations of motion, ~\eqref{eq:one} and \eqref{eq:two}, are integrated forward in time from $t = 0$ to $t = \tF$ with the current protocol of spatiotemporal activity $\alpha^{i}(\mathbf{r}, t)$ (where $i$ is the current iteration) and fixed initial conditions, $\mathbf{Q}(\mathbf{r}, 0)$ and $\mathbf{u}(\mathbf{r}, 0)$. 
    \item  \textbf{Step 2:} The adjoint equations, ~\eqref{eq:adjoint}, are integrated backward in time from $t = \tF$ to $t = 0$ with the initial condition, $\boldsymbol{\psi}(\mathbf{r}, \tF) = 0$ and $\boldsymbol{\nu}(\mathbf{r}, \tF) = 0$.
    \item \textbf{Step 3:} The control protocol is updated via gradient descent, $\alpha^{i+1} = \alpha^{i} - \Delta {\delta \mathcal{J}}/{\delta \alpha} $, to minimize the cost function. 
    \item \textbf{Step 4:} The calculated control protocol is clipped such that $\alpha(\mathbf{r}, t) \geq 0$, to satisfy the extensile constraint. 
\end{itemize}
Iterations are continued until the gradient $\delta \mathcal{J}/{\delta \alpha}$ falls below a user-defined tolerance, which we set to $10^{-4}$. We employ Armijo backtracking \cite{Borzi2011} to adaptively choose the step-size $\Delta$ for gradient descent and to ensure convergence of the DAL algorithm.

\section{Controllability}

\label{sec:controllability}

Let us consider the problem we want to solve in control theory. If we define $\mathbf{X}=\left(\begin{array}{c} {Q}_{\text xx} \\ {Q}_{\text xy}\\ \end{array}\right)$, we seek to solve the set of nonlinear partial differential equations $\frac{\partial \mathbf{X}}{\partial t}=H\left[\mathbf{X},\alpha \right]$ for the control solution $\alpha(\mathbf{r},t)$, subject to a given initial condition $X(\mathbf{r},0)$, and a boundary condition in time $X(\mathbf{r},t_{\text{F}})$, which is the target state with $\{0,t_{\text F}\}$ as the control window. A particular dynamical system is considered controllable if we can demonstrate the existence of a solution to the above problem.  When the dynamics is nonlinear, demonstrations of controllability have been limited to a few simple systems where the nonlinearities have special properties.  What we do instead is consider the controllability of Eqs. \ref{eq:one} -\ref{eq:two}  when linearized about the unstable fixed point of a uniformly aligned state.  Demonstrating controllability of the homogeneous fixed point tells us that at short enough length scales, we will be able to drive the system to desired values of the dynamical fields, which can be thought of as different fixed points in the continuous space of fixed points associated with translational symmetry and broken rotational symmetry characteristic of our system.

Linearizing our theory using  $Q_{\text ij} = Q_{\text ij}^{0} + \delta Q_{\text ij}(\boldsymbol{r}, t)$ and, $u_{\text i} = u_{\text i}^{0} + \delta u_{\text i}(\boldsymbol{r}, t)$ and introducing the Fourier transform, $\Tilde{x}(q, t) = \int d\boldsymbol{r}e^{i\boldsymbol{q\cdot r}}x(\boldsymbol{r}, t)$, we obtain
\begin{align}
    \nabla^{2} \delta \mathbf{u} - \nabla \delta P - \xi \delta \mathbf{u} = \mathbf{Q^{0}}\cdot \nabla \delta \alpha + \alpha \nabla \cdot \delta \mathbf{Q}
\end{align}

Pressure and velocity in Fourier space are given by
\begin{align}
    & \delta \Tilde{P} =  \frac{\delta \Tilde{\alpha}}{q^{2}}\left[ q_\text {i} q_\text{j} Q_{ \text ij}^{0}\right] + 
    \frac{\alpha}{q^{2}}\left[ q_\text{i} q_\text{j} \delta \Tilde{Q_{\text ij}}\right] \\
    & \delta \Tilde{u}_{i} = \frac{-1}{(q^{2} + \xi)}( i q_{\text i}\delta \Tilde{P} + Q_{\text ij}^{0} i q_{\text j} \delta \Tilde{\alpha} + \alpha i q_{\text j} \delta \Tilde{Q_{\text ji}})
\end{align}

Eq. \ref{eq:one} can be written in Fourier space in scalar form as 

\begin{widetext}

\begin{align}
    & \partial_{t} \delta \Tilde{Q}_{xx} = -(i u_x^0 q_x + i u_y^0 q_y) \delta \Tilde{Q}_{xx}-Q_{xy}^{0}(i q_y \delta\Tilde{ u}_{x} - i q_x \delta \Tilde{u}_y) + \lambda i q_x \delta \Tilde{u}_x  +  (a_2 - a_4 s^2) \delta \Tilde{Q}_{xx} - K q^{2} \delta \Tilde{Q}_{xx} \\
    & \partial_{t} \delta\Tilde{Q}_{xy} = -(i u_x^0 q_x + i u_y^0 q_y) \delta \Tilde{Q}_{xy} + Q_{xx}^{0}(iq_y \delta \Tilde{u}_x - i q_x \Tilde{u}_y ) + \frac{\lambda}{2} (i q_x \delta\Tilde{u}_y + i q_y \delta\Tilde{u}_x) + (a_2 - a_4 s^2 )\delta \Tilde{Q}_{xy} - K q^{2} \delta \Tilde{Q}_{xy}
\end{align}
    
\end{widetext}


Here $a_2, a_4> 0$. Using fixed points $\{u_x^0, u_y^0, Q_{xx}^{0}, Q_{xy}^{0} = 0, 0, \frac{1}{2}, 0 \}$, we write the linearized equation of motion in Fourier space as
\begin{widetext}

\begin{equation}
\begin{aligned}
\partial_t\begin{pmatrix}
\delta \Tilde{Q}_{xx} \\
\delta \Tilde{Q}_{xy} 
\end{pmatrix} = &
\begin{pmatrix}
  \lambda \frac{q_x^4}{ q^2(q^2 + \xi)}2\alpha  + a_2 - a_4 s^2 -Kq^{2}  &  \lambda \alpha\frac{q_x q_y }{q^2 + \xi}(\frac{2 q_x^2}{q^2} + 1)  \\
\frac{q_x q_y}{(q^2 + \xi)}\alpha + \alpha \lambda \frac{q_x^2 - q_y^2}{q^2}\frac{q_x q_y}{q^2 + \xi}& -\frac{(q_x^2 - q_y^2)}{2(q^2 + \xi)}\alpha + \frac{\alpha \lambda}{2(q^2 + \xi)}(\frac{4 q_x^2 q_y^2}{q^2} + q^2)  + a_2 - a_4 s^2  -Kq^{2}
\end{pmatrix} \\
& \times \begin{pmatrix}
\delta \Tilde{Q}_{xx} \\
\delta \Tilde{Q}_{xy} 
\end{pmatrix} 
+ B_q \delta \Tilde{\alpha}
\end{aligned}
\end{equation}

\end{widetext}
where

$$ B_{\text q} = 
\left(\begin{array}{c}
 \lambda \frac{q_x^4}{q^2 (q^2 + \xi)}\\[10pt]
 \left(1 +\lambda \frac{q_x^2 - q_y^2}{q^2}\right) \frac{q_x q_y}{2(q^2 + \xi)}
\end{array}\right).
$$
Our linearized theory is then of the form 
$$
\partial_{t} X_{\text q}(t)=A_{\text q} X_{\text q}(t)+B_{\text q} \delta\alpha_{\text q}(t)
$$

One can readily establish that this linear system has a solution to the boundary value control problem when  the controllability matrix
$$
C=\left[\begin{array}{lllll}
B_{\text q} & A_{\text q} B_{\text q}
\end{array}\right]
$$
is of rank 2 \cite{Brunton2022}.

Computing the column vector of controllability matrix

$$ A_{\text{q}}B_{\text q} = 
\left(\begin{array}{c}
2\alpha\lambda^2 \frac{q_x^8}{q^4 (q^2 + \xi)^2} + (a_2 - a_4 s^2 - Kq^2)\lambda \frac{q_x^4}{q^2 (q^2 + \xi)} \\
+ \frac{\lambda \alpha} {2} \frac{q_x^2 q_y^2}{(q^2 + \xi)^2}\left( \frac{2 q_x^2}{q^2} + 1\right) \left( 1 + \lambda \frac{q_x^2 - q_y^2}{q^2}\right)

\\[15pt]

\left( \frac{q_x q_y}{q^2 + \xi} \alpha + \alpha \lambda \frac{q_x^2 - q_y^2}{q^2} \frac{q_x q_y}{q^2 + \xi}\right)\lambda \frac{q_x^4}{q^2 (q^2 + \xi) } \\
+ \left(
\frac{\alpha \lambda}{2(q^2 + \xi)} \left( \frac{4 q_x^2 q_y^2}{q^2} + q^2\right) - \frac{q_x^2 - q_y^2}{2(q^2 + \xi)} \right) \left( 1 + \lambda \frac{q_x^2 - q_y^2}{q^2}\right)\frac{q_x q_y}{2 (q^2 + \xi)} \\
+ (a_2 - a_4 s^2 - K q^2)\left( 1 + \lambda \frac{q_x^2 - q_y^2}{q^2}\right)\frac{q_x q_y}{2 (q^2 + \xi)}

\end{array}\right).
$$

We can see that when gradients in both $x$ and $y$ direction are present, the linearized theory is controllable.

In a special case when $q_y = 0$, the controllability matrix becomes
\begin{equation}
  C =   \begin{pmatrix}
         \lambda \frac{q_x^2}{q_x^2 + \xi} & 2 \alpha \lambda^2 \frac{q_x^4 }{ (q_x^2 + \xi)^2} \\  & + \lambda(a_2 - a_4 s^2 - Kq^2) \frac{q_x^2}{q_x^2 + \xi}  \\[10pt]
         0 & 0
        \end{pmatrix} 
\end{equation}
The column vectors of this matrix are not linearly independent, hence the linear theory becomes uncontrollable when gradients perpendicular to the direction of broken symmetry vanish. Thus, we see that spatial gradients orthogonal to the direction of the local
alignment are critical to obtaining control solutions for an active nematic fluid. 

\begin{widetext}

\section{Additional figures.}
\begin{figure*}[h!]
  \includegraphics[width=\textwidth]{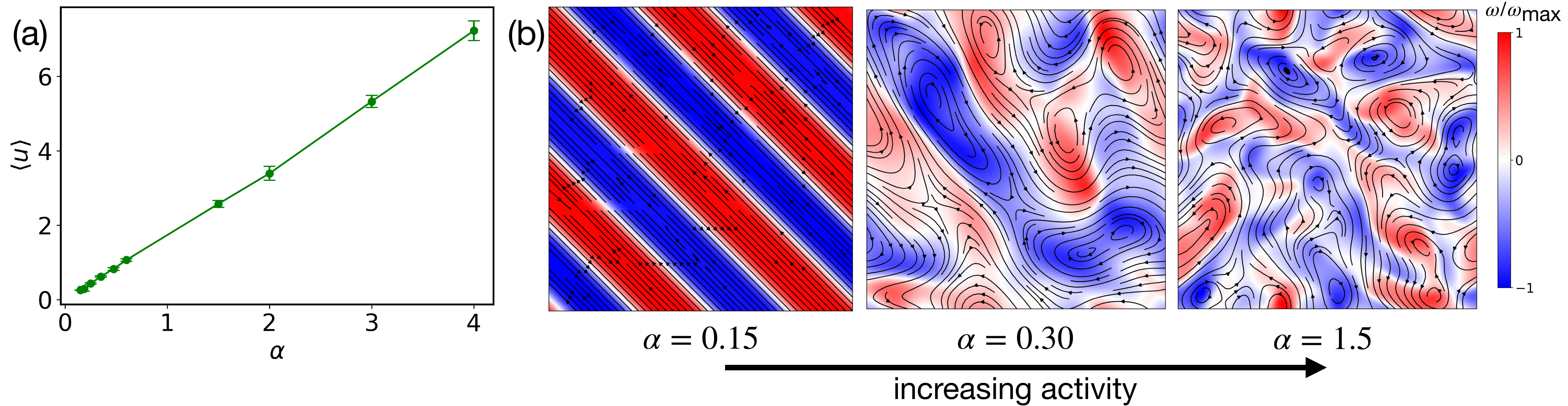}
  \caption{ \textbf{Additional properties of the uncontrolled bulk active nematic as a function of activity.} (a) Average velocity. (b) Vorticity (colormap) and velocity field (arrows). 
}
  \label{fig:umean}
\end{figure*}

\begin{figure*}[h!]
  \includegraphics[width=\textwidth]{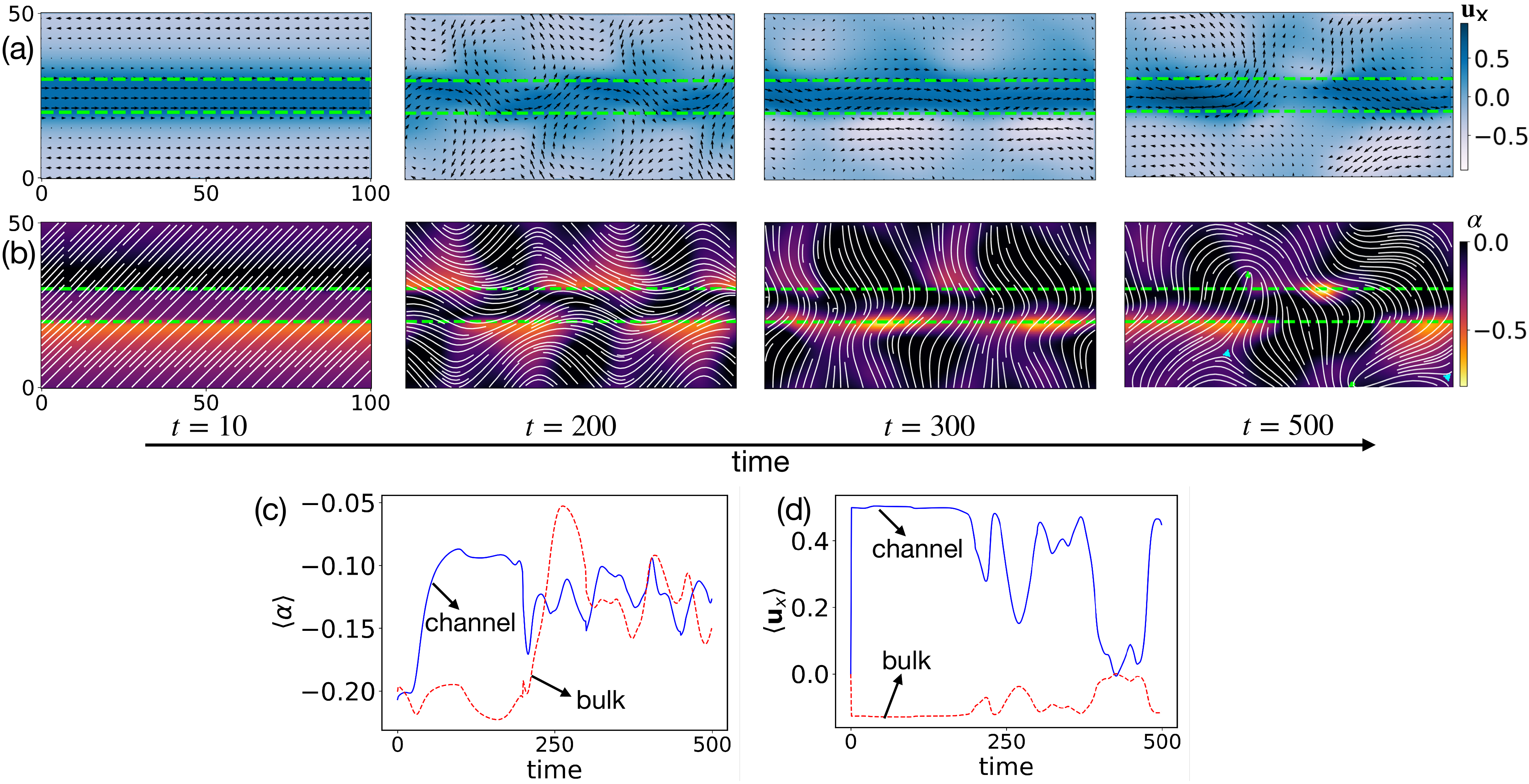}
  \caption{ \textbf{Coherent flow through a `channel' without confining boundaries.} The target state has uniform flow along the $\hat{x}$ direction with magnitude $\Utarget=0.5$ within the strip enclosed by green dashed lines.  \textbf{(a)} The velocity field (arrows) and magnitude $u$ (color map) are shown at indicated time points for a representative trajectory under the optimal control protocol. \textbf{(b)} The activity field $\alpha$ (color map) and nematic director field (lines) are shown at the same time points. 
 The initial state has the nematic uniformly aligned along the $\pi/4$ axis. 
\textbf{(c)} The flow speed  $u_x$ \textbf{(c)} and  the activity \textbf{(d)} averaged over the channel and bulk regions as a function of time.
 The parameter values for this solution were $\{J_1, J_2, J_3, \zeta, \wU, \wQ, \alpha_0, \tF\} = \{1.0, 1.0, 0.1, 0.1, 30.0, 0, 0.3, 500.0\}$.  A video of this trajectory is provided in supplemental movie S5 \cite{SIref}.
}
  \label{fig:flowhigh}
\end{figure*}

\begin{figure*}[h!]
  \includegraphics[width=\textwidth]{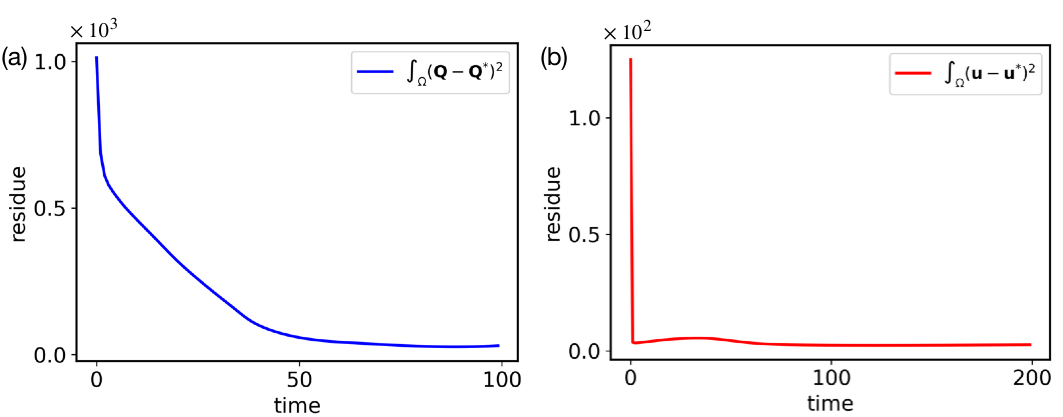}
  \caption{ \textbf{Residues measured between system state and target state.} 
  Deviation of the control trajectory from the target trajectory as a function of time for the solutions shown in \textbf{(a)} Fig. \ref{fig:ordernematic} and \textbf{(b)} Fig. \ref{fig:coherentflow}.
}
  \label{fig:residue}
\end{figure*}

\end{widetext}
\section{Movie descriptions}
\begin{itemize}
    \item \textbf{Movie S1}: Aligning the director field in a square subdomain (Fig. 3 main text). The left panel shows the nematic order parameter $s$ (color map) and director field $\mathbf{Q}$ (lines). The right panel shows the activity field $\alpha$, with a logarithmic scale colorbar.
    \item \textbf{Movie S2}: Coherent flow through a channel without confining boundaries (Fig. 4 main text) for $L_y = 50$. The top panel shows the x-velocity $\mathbf{u}_{\mathbf{x}}$ (color map) and director field $\mathbf{Q}$ (lines). The bottom panel shows the activity field $\alpha$, with a linear scale colorbar.
    \item \textbf{Movie S3}: Coherent flow through a channel without confining boundaries (Fig. 4 main text) for $L_y = 100$. The top panel shows the x-velocity $\mathbf{u}_{\mathbf{x}}$ (color map) and director field $\mathbf{Q}$ (lines). The bottom panel shows the activity field $\alpha$, with a linear scale colorbar.
    \item \textbf{Movie S4}: Couette flow in a channel without a moving boundary (Fig. 5 main text). The top panel shows the x-velocity $\mathbf{u}_{\mathbf{x}}$ (color map) and director field $\mathbf{Q}$ (lines). The bottom panel shows the activity field $\alpha$, with a linear scale colorbar.
    \item \textbf{Movie S5}: Coherent flow through a channel without confining boundaries  with high activity baseline (Fig. \ref{fig:flowhigh}). The top panel shows the x-velocity $\mathbf{u}_{\mathbf{x}}$ (color map) and director field $\mathbf{Q}$ (lines). The bottom panel shows the activity field $\alpha$, with a linear scale colorbar.
\end{itemize}


\bibliography{fromjabref, ControlPapersForDOE}

\end{document}